# CRITICAL TASK RE-ASSIGNMENT UNDER HYBRID SCHEDULING APPROACH IN MULTIPROCESSOR REAL- TIME SYSTEMS

Gopalakrishnan T.R. Nair[1], Christy A. Persya[2]

[1]Saudi Aramco Endowed Chair - Technology, Prince Mohammad University, Vice President, RIIC, DS Institutions, India

[2]Research Associate, Real Time Systems Group, RIIC, Sr.Lecturer, TOCE, India.

[1]trgnair@gmail.com, [2]christypersya@gmail.com

**ABSTRACT**

Embedded hard real time systems require substantial amount of emergency processing power for the management of large scale systems like a nuclear power plant under the threat of an earth quake or a future transport systems under a peril. In order to meet a fully coordinated supervisory control of multiple domains of a large scale system, it requires the scenario of engaging multiprocessor real time design. There are various types of scheduling schemes existing for meeting the critical task assignment in multiple processor environments and it requires the tracking of faulty conditions of the subsystem to avoid system underperformance from failure patterns. Hybrid scheduling usually engages a combined scheduling philosophy comprising of a static scheduling of a set of tasks and a highly pre-emptive scheduling for another set of tasks in different situations of process control. There are instances where highly critical tasks need to be introduced at a least expected catastrophe and it cannot be ensured to meet all deadline in selected processors because of the arrival pattern of such tasks and they bear low tolerance of time to meet the required target. In such circumstances an effective switching of processors for this set of task is feasible and we describe a method to achieve this effectively.

**KEY WORDS**
Hybrid Scheduling, Embedded, Pre-emptive

## 1. INTRODUCTION

Reliable embedded systems play a leading role in the current era applications like avionics, nuclear power plant etc. In order to meet several requirements, fixed scheduling and flexible scheduling are balanced through stringent deadlines. Many studies have sought to develop a feasible hybrid task schedule and fault tolerant schemes. Feasible hybrid schedule can be achieved by using Rate Monotonic and Earliest Deadline First (EDF) scheduling algorithm. Both algorithms have its own pros and cons as discussed by Jukka Maki [2].

The critical task reassignment strategy has become a challenging proposition in multiprocessor real time system for managing large scale operations. Inevitably, real time system calls for timely response for certain signals of low probability to avoid catastrophic situation and lossy business situations created by sudden damages in subsystems like in earth quake or an explosion in a sensitive production plant.

In a large scale system, there is a definite proportion of task which could be enlisted through its parametric observations which justifies them to be included in the fixed priority segment. Since the system being a large ordered system, there are going to be few tasks that have got rarest occurrence but have high catastrophic value like a brake failure of an automotive system or a fuel leakage of the gas turbine in aircraft or eminence of a catastrophic radiation in nuclear installations or the occurrence of an earth quake in real time which rarely happens. So probabilistically a hybrid schedulers packed the tasks nicely in such a way that binding of tasks to each one of the processors are most efficiently handled and time windows are almost fully loaded. So it requires a high priority super scheduler to systematically stop one or several tasks in one or more than one processor to achieve the implementation of catastrophic controls. We introduce such a theme and discuss some of the terms and conditions of implementing this and deal with the efficiency and drawbacks through the simulation results.

The rest of the paper is structured as follows: the hybrid scheduler model is discussed in section 2.section 3 deals with the research background. The architecture of the new scheduler and the priority alter protocol is discussed in section 4.section 5 presents performance evaluation. Finally, in section 6, some concluding remarks are made.

## 2. SYSTEM MODEL

In this section, we first present the scheduler model, followed by task model with some definitions which are necessary to explain the scheduling algorithm.

### 2.1 Hybrid Scheduler Model

The Hybrid scheduler is the combination of the static and dynamic scheduling pattern. Two phase architectural model is used here. First, identify the group of tasks to be scheduled together with an offline or fixed scheduler (periodic tasks). Second, primary backup can be used for dynamic priority tasks with effective communication mechanism with processors to synchronize. The backup copies can be overlapped which can reduce the number of backup processors [10][11]. An m:q ratio of primary and backup processors is used in the architecture. Segmented shared memory can be used to have the advantage of simultaneous sharing of memory by multiple programs [12][13].

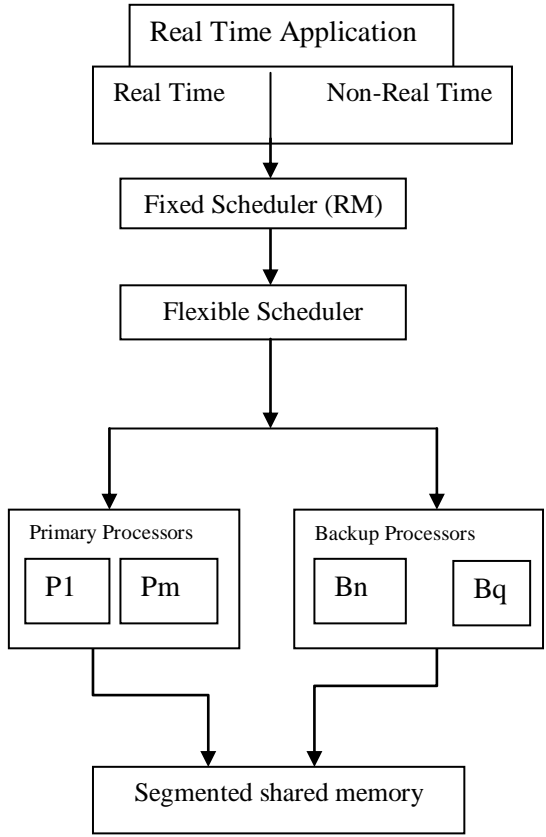

Fig 1: Temporal positioning of scheduling scheme

## 2.2 Task Model

To define a scheduling problem, it is required to specify a set of n tasks $\Gamma = \{\tau_1,...,\tau_n\}$, and the temporal sequence of task executions. We consider tasks characterized by an execution time C, an inter-arrival time T and a relative deadline D describing the temporal interval within which the task has to be executed. Each task $\tau_i$ is a sequence of job $J_{i,j}$, which means that the j-th job of $\tau_i$ is released at time instant $(j-1)T_i$ and it must finish its execution by time instant $(j-1)T_i + D_i$. Classical task model denotes $\tau_i$ by $(C_i, T_i, D_i)$, where Release time is the instant of time at which the job becomes available for execution. Deadline is the instant of time by which its execution is required to be completed. Execution time is the computation time. Period is the interval between the occurrences of the task. Priority is the right to precede others in order, rank, privilege, etc.

## 3. RESEARCH BACKGROUND

Here, first we discuss the existing work on Hybrid scheduling on uniprocessor with periodic or aperiodic/sporadic tasks and then highlight the limitations which form the motivation for our work.

### 3.1 Related Work

Many application scheduling problems for critical tasks are found to be NP complete i.e., it is believed that there is no optimal solution for scheduling.

Researchers have proposed Hybrid scheme for Hard, Soft and non-real time tasks [1] where the two level architecture of hierarchical scheduling scheme was used and proved that Hybrid scheme has higher synthetic performance than pure EDF and pure RM scheme. This was analyzed on a single processor and the results motivated the need to schedule for multi processors.

In [2].an analysis for responsiveness of dynamic tasks under scheduling schemes like hybrid, static and dynamic models have been developed and the results guaranteed that the hybrid scheduling model simplifies the design tradeoffs of selecting the scheduling model.

In [3], the design and development of new CPU scheduling algorithms (the Hybrid scheduling algorithm and the Dual Queue scheduling algorithm) has been designed to optimize resource utilization.

In [14], a hybrid scheduling approach for real-time systems on homogeneous multi-core architectures have been developed and allows the real-time applications to run with non-real-time applications concurrently and supports the parallelism between the tasks within an application efficiently.

In [4], the dynamic scheduling algorithm with PB fault-tolerant algorithm is mapped to improve the guarantee ratio, which is the percentage of tasks arrived in the system whose deadlines are met.

A PB based algorithm with flexible backup overloading in Distance Myopic has to be proposed for fault tolerant dynamic scheduling of real tasks in multiprocessor systems, which we call myopic backup overloading. Backup overloading is a process of allocating a single backup slot for more than one task. So when a single task failed in a processor, it can be switched to the backup processor when the other tasks are running on their scheduled primary processor.

### 3.2 Problem definitions

This paper considers the Task Reassignment Problem (T-RA-P) with the following scenario. The system consists of a set of heterogeneous processors (m) having different memory and processing resources, where each processor is tightly packed with real time tasks. The highly critical catastrophic set of tasks occurs at an unusual pattern at unpredictable time, failure to deal with it, will lead to a major catastrophic scene. When this task is arrived, all processors in the homo/heterogeneous multiprocessor systems are more or less fully loaded with their scheduled tasks or accommodating a task set extra in normal scheduled sequence is cumbersome. All these scheduled tasks are hard real time tasks, which cannot be pre-empted by the arrived critical task. The scheduler in each processor communicates through the message passing system.

Here, if tasks are unpredictable, they cannot derive accurate replacement rules. It will run the server under the constant utilization rules, giving $qu_i$ units of budget every q time units. The scheduling quantum q is a key parameter.

The paper deals with the possible task reassignment policies that can be made such that no catastrophic situation takes place or the chance for it is minimized and how the scheduler handles the situation such that the stability condition of processor and the tasks persists.

## 4. ARCHITECTURE OF THE SCHEDULER

### 4.1 Improved Scheduler

This section presents the design and analysis of our scheduling scheme, we take (1) All the tasks with EDF algorithm $T_{EDF}$ and all the tasks with RM algorithm $T_{RM}$ (2) Applications $T_{EDF}$ and $T_{RM}$ are executed by constant utilization server $U_{EDF}$ and $U_{RM}$ (3) The size of the constant utilization server of $T_{EDF}$, α is 0<α<1.The size of the constant utilization server of $T_{RM}$, β is 0<β<1.

Theory of Super Schedulers:

Super Schedulers are the highest prioritized dispatcher of jobs. Hybrid schedulers are inbuilt into the super schedulers (inner region).The rest of the super scheduler is covered with the catastrophic scheduler which purely acts when an unusual catastrophic critical set of tasks arrive. All together the entire scheduler system will be working as hybrid scheduler, which is triggered by the super scheduler.

The hybrid scheduler formalism is used for critical task reassignment. It is completely different from the interrupt service routine and handler.

The architecture of Super Scheduler is presented in Fig 2.

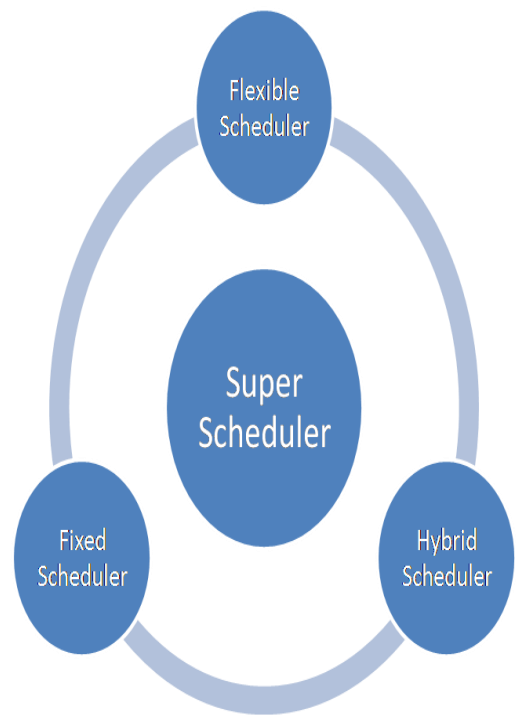

Fig 2: Super scheduler architecture

In the major cycle, when the catastrophic critical task arrives, the scheduler automatically performs the context switching of Hybrid schedulers by the super scheduler. So the priority of the currently running high priority tasks are altered with the arrived catastrophic critical task. The priority altered critical tasks are reassigned by the super scheduler. When the critical task completes its process, then the context switching of altered tasks from the stack starts execution. Here, few low priority tasks cannot meet its deadline and will be discarded, since they don't cause any major disaster. There is no additional execution time of the systems when using the super scheduler as well in non-critical situations.

As a reduced order, we consider the system with fourtasks:T1=(20,25,150),T2=(40,10,50),T3=(60,50,200) and T4=(50,30,180). Suppose that in addition to the critical tasks that has execution time 60 and is released at 60. We call this task CT1.Figure 3 shows the schedule and execution of super scheduler.

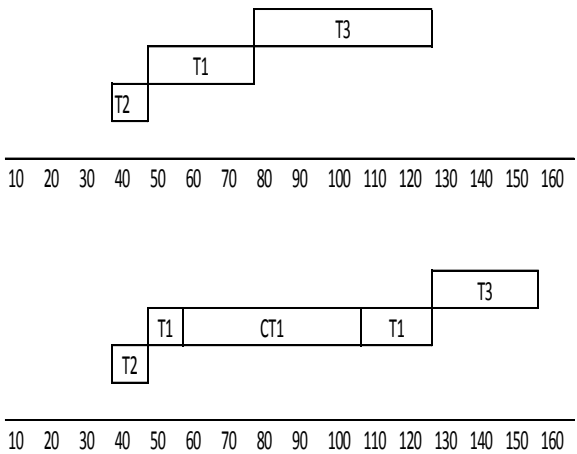

Fig 3: Schedule and execution of super scheduler

1. Initially, the super scheduler is suspended because the critical task queue is empty. When CT1 arrives at 60, the super scheduler resumes because the execution of the last 15 units of T1 can be pre empted until time 110.

2. At time 110, CT1 completes. The super scheduler is suspended. The task T1 resumes and executes to completion on time.

3. For as long as there is no job in the critical queue, the normal schedule execute on the EDF basis.

The super scheduler starts its execution at a point where the next cycle begins by altering the priorities of the previously scheduled tasks. It re schedules the execution of the tasks whenever one of the following events occurs: a major catastrophic critical task occurs - an estimated duration becomes shorter or longer - modification of resource, activity, priority, duration, etc. - resource breakdowns - activities and links are discarded - resume after a crash.

The Advantages of our scheduler:

The innovative features of this task reassignment approach certainly set it apart from other scheduling engines, but, more importantly, they promise to deliver to the embedded industry the efficiency and flexibility that users have a right to expect in an automated system with catastrophe management concept

- Fastest time-to-respond and flexible to redesign
- No further need to spend valuable time developing complex scheduling technologies
- Resources will stay focused on system
- It can take advantage of hybrid scheduling, efficiency and performance improves.

### 4.2 Task Reassignment Policies

The system can always reassign the tasks when the high prioritized critical task arrives. This reassignment will be initiated and executed by the catastrophic scheduler. The assumptions followed in this paper are

**Assumption 1**: All the real time tasks are independent, periodic, and pre-emptive and their relative deadlines are equal to its period.

**Assumption 2:** The relative deadline and the computation time of every real time task are known when it is in ready state.

**Assumption 3:** All the critical tasks are re schedulable by the algorithm when the algorithm always produces a feasible schedule.

All Theorems and statements made in this paper are related and interdependent.

The facts to be considered during reassignment are:

**Lemma 1:** A scheduling algorithm can feasibly schedule any set of periodic tasks on a processor if the total utilization of the tasks is equal to or less than the schedulable utilization of the algorithm.

The disadvantage of the dynamic scheduling EDF algorithm is when a hard real time task which has already missed its deadline has a highest priority than a task whose deadline is still in the future, a good overrun management strategy is vital to prevent this kind of situation. An overrun management strategy is to reschedule the late task a lower priority than the tasks that are not late and more critical.

**Hypothesis**: In hybrid scheduling scheme, a critical task $CT_i$ arrives in a system in an unusual manner. All pre-emptive periodic tasks have definite chance to be pushed to postpone state by the super scheduler by transferring the priorities of current running task $T_i$ with critical task $CT_i$.

**Proof.** The system is not feasible if its total utilization is greater than 1. EDF algorithm is optimal, i.e., it can produce a feasible schedule of any feasible system. Hence, it can be stated that the EDF algorithm can surely produce a feasible schedule of any system with total utilization less than 1. Since priorities are based on deadlines $d_i$, when a most critical task arrives, priorities are explicitly transferred to the arrived task $CT_i$ and remaining routine task $T_i$. So the least prioritized task may miss its deadline and discarded which may not cause major disaster in safety critical applications.

**Lemma 2:** The system is said to be overloaded when the jobs presented to the scheduler cannot be feasibly scheduled even by supernatural scheduler.

In the above example after the reassignment of tasks, the system is overloaded so the task T4 misses its deadline which is least prioritized.

### 4.3 Priority Alter Protocol

The Priority alter protocol is given as

---

1. **The most critical task $CT_i$ is given the highest priority by the super scheduler and is assigned to the processor. This is predesigned for the most uncommon catastrophe.**

2. **Suppose $CT_i$ suspends/interrupts one or more tasks. Then, it alters the priority of the highest priority task T that is currently running, to the newly arrived most critical task $CT_i$ under super scheduler**

3. **A task $T_i$ can preempt another task $T_j$ and their priority precedence will take the order as**

   $CT_i > T_i > T_k > \ldots\ldots\ldots > T_j$.

---

The key properties of the priority alter protocol are as follows:

P1: The priority alter protocol prevents catastrophic deadline miss.

P2: It alters the priority between the new major catastrophic tasks with other scheduled tasks and performs a reschedule of tasks.

## 5. PERFORMANCE EVALUATION

**Pseudo code for a Super Scheduler**

The following is the pseudo code for a super scheduler we discussed which schedules all critical periodic, aperiodic and sporadic tasks. It is assumed that the precomputed schedule for periodic tasks by hybrid scheduler is stored in a schedule table

**Super – Scheduler ( ) {**

Current – task T = Schedule – table [K];

K = K+1 ;

K = K mod N ;  // N is the total no of schedule

                 // tasks in the schedule

Activates priority alter protocol

Dispatch – current – Task (T);

Schedule – critical Task ( );

Reschedule – Suspended Tasks ( );

Idle ( );

}

The super scheduler routine super – Scheduler ( ) is activated at the entry of every severe critical task. Then the priority of the current task is altered with the critical task and the current task is suspended and rescheduled to be run in the next frame by invoking the routine super scheduler ( ) . If some of the tasks could not complete its execution, it is discarded by the scheduler based priority based format.

Super scheduler does not hold the control of a processor full time because catastrophic tasks will not run every time. They run only in emergency situations like brake failure of an automotive system or a fuel leakage of the gas turbine in aircraft or emergence of a catastrophic radiation in nuclear installations or the occurrence of a earth quake in real time which probably never happens.

As per the theorem discussed by Liu & Leyland in [9] , (93)

**Theorem:**

No on-line scheduling algorithm can achieve a competitive factor greater than 0.25 when the system is overloaded.

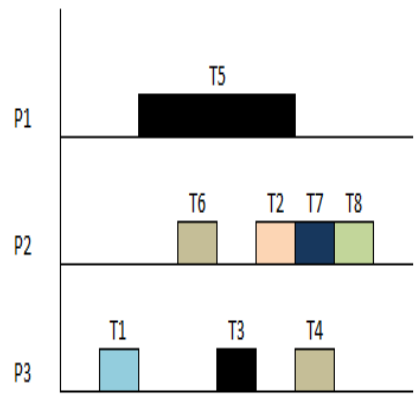

Fig 4: Execution of scheduled tasks by hybrid scheduler

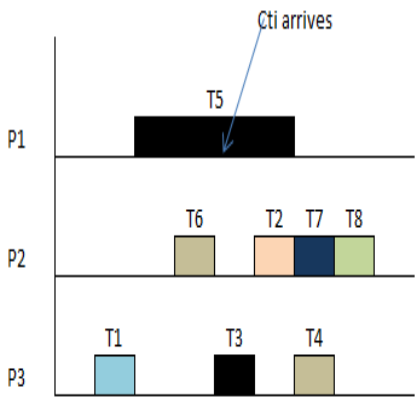

Fig 5: When major critical task CTi arrives

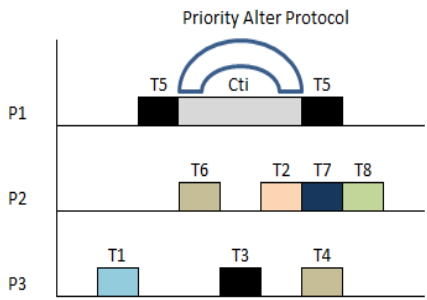

Fig 6: Stability of processor even after super scheduler wakes with priority alter protocol

Fig 4 explains the execution of hybrid scheduler of tasks by three processors. Hybrid scheduler takes the advantage of rate monotonic and earliest deadline first algorithm. In fig 5 critical task **$CT_i$** interrupts task T5. Super Scheduler is activated and it initiates the priority alter protocol. Critical task **$CT_i$** is assigned to the processor and other tasks are rescheduled or reassigned to the processors as shown in fig 6. The stability of the system is also maintained.

Through experimental results and the theorems stated above, it is proved that the instance where highly critical task cannot meet its deadline with any hybrid scheduler. These tasks are ensured to meet its deadline with the new super scheduler approach.

As stated in [1], the important performance criterion is the deadline missing rate, which is defined as the ratio of the number of the real-time task instances having missed their deadlines to the number of all real-time task instances.

The stability condition of processors can be derived as follows:

Let the success rate of processor before the arrival of unpredictable task is X and the success rate of processor after scheduling the unpredictable task by super scheduler be Y.

So, the overall success rate of feasible schedule will be X < Y.

The stability condition of the processor schedule even after the arrival of unpredictable task can be computed as

$$T(X) - \frac{Miss(n)}{N} > 0.7$$

Where,

T(X) – No of tasks waited for scheduling after the arrival of unpredictable tasks
Miss (n) -  No of least priority tasks missed its deadline
N - Total no of tasks scheduled to the processor

So, if the processor utilization factor is greater than 0.7 even after the arrival of unpredictable tasks, then the stability condition of processors can be maintained.

When the unexpected critical task arrives, the super scheduler takes control of the entire system and alters the priority of the running task with the newly arrived critical task. In this way, the super scheduler changes the predetermined scenario and reschedules the tasks. So in this case because of task reassignment and priority alteration, few tasks which are least prioritized will miss its deadline. So the task miss rate is assumed as 0.3.i.e., maximum miss rate can be 30 % of tasks. So when miss rate is 30 % then the success rate of the tasks with processor utilization can be determined as 0.7. The stability condition of the system persists if the success rate is more than 0.7.

On the whole, the performance of the super scheduling scheme presented in this paper is higher than the hybrid alone scheme.

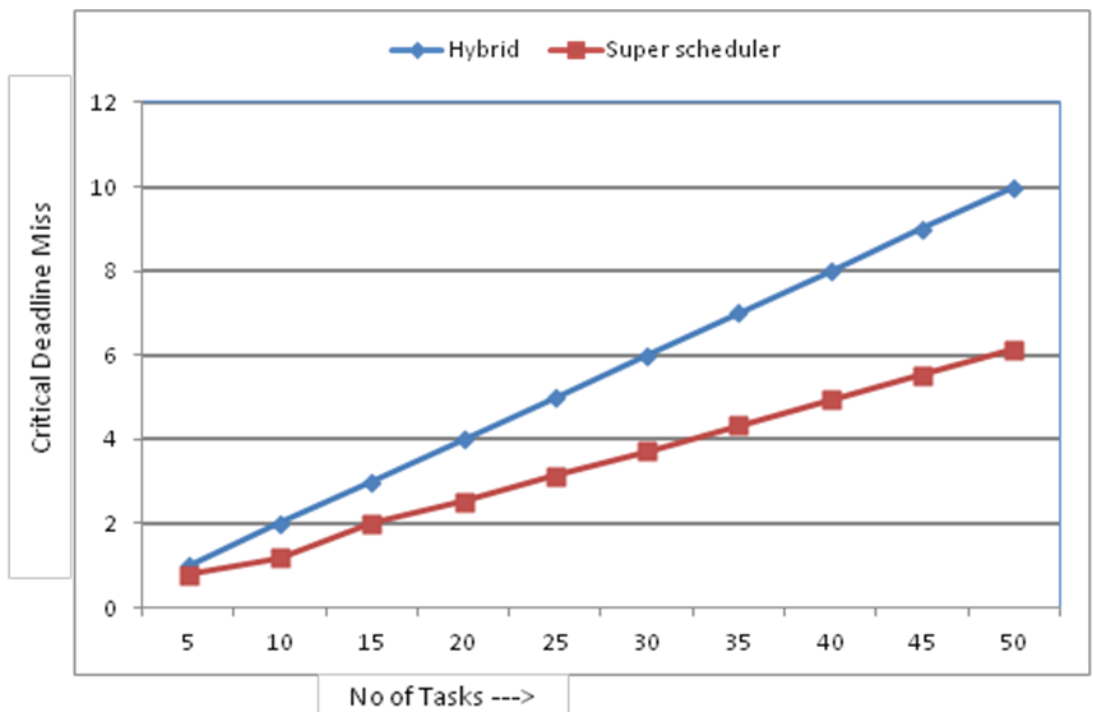

Fig 6: Variance of critical deadline miss

## 6. CONCLUSION

While hybrid scheduling approach is viable, the adaption of super scheduler approach offers greater advantages, most notably meeting the deadlines of all critical tasks which may arise at external catastrophe to plants.

In this paper, we present a new approach for handling the most critical task which arrives at an unusual pattern leading to great loss. In order to handle this situation, a new scheduler which has the inherited features of hybrid scheduler called super scheduler is introduced and the results of phase 1 study were presented.

It is also observed that the stability condition of processors will persist even after the appearance of unusual critical task introduced by the super scheduler.

Future Research can focus on implementing task reassignment algorithm and can derive the system stability conditions along with the impact of critical task in multi core system.